\documentclass{article}

\usepackage{PRIMEarxiv}

\usepackage[utf8]{inputenc} 
\usepackage[T1]{fontenc}    
\usepackage{hyperref}       
\usepackage{url}            
\usepackage{booktabs}       
\usepackage{amsfonts}       
\usepackage{nicefrac}       
\usepackage{microtype}      
\usepackage{lipsum}
\usepackage{fancyhdr}       
\usepackage{graphicx}       

\usepackage{amsmath}

\DeclareMathOperator*{\argmin}{arg\,min}

\graphicspath{{media/}}     

\title{Competitive learning to generate sparse representations for associative memory
}

\author{
  Luis Sacouto, Andreas Wichert \\
  INESC-ID \& Department of Computer Science and Engineering \\
  Higher Technical Institute, University of Lisbon (IST-UL) \\
  Lisbon, Portugal\\
  \texttt{\{luis.sa.couto, andreas.wichert\}@tecnico.ulisboa.pt} \\
}

\begin{document}
\maketitle

\begin{abstract}
One of the most well established brain principles, hebbian learning, has led to the theoretical concept of neural assemblies. Based on it, many interesting brain theories have spawned. 
Palm's work implements this concept through binary associative memory, in a model that not only has a wide cognitive explanatory power but also makes neuroscientific predictions.
Yet, associative memory can only work with logarithmic sparse representations, which makes it extremely difficult to apply the model to real data.
We propose a biologically plausible network that encodes images into codes that are suitable for associative memory. It is organized into groups of neurons that specialize on local receptive fields, and learn through a competitive scheme.
After conducting auto- and hetero-association experiments on two visual data sets, we can conclude that our network not only beats sparse coding baselines, but also that it comes close to the performance achieved using optimal random codes.
\end{abstract}


\section{Introduction}
With the advent of Deep Learning, artificial intelligence systems are getting better at solving perceptual tasks such as visual pattern recognition. Although these systems have achieved tremendous success in many tasks they still lack several desirable properties that biological brains possess, among which are behavior flexibility and energy efficiency \cite{zador_toward_2022}. This makes computational neuroscience a key field for basic research, not only from the ``understanding the brain'' perspective, but also in trying to develop intelligent systems that "work".

The intersection between the two views is not as popular as each view by itself, partially because no technique seems to achieve Deep Learning results on the most common machine learning benchmarks, but also because the brain, as an intelligent system, is still largely a mystery for modern science.

Despite this discouraging fact, there are some findings about the brain which are well grounded and widely accepted. One of the most important is the idea of hebbian learning, originally proposed by Hebb \cite{hebb_organization_1949}, and later supported by several studies \cite{stuart_active_1994,citri_synaptic_2008,oreilly_six_1998,szatmary_spike-timing_2010,barak_working_2014,mi_synaptic_2017,fiebig_spiking_2017}. The principle states that when two neurons fire together, the synapse between them will strengthen, thus making them more correlated. 

Building upon this idea, many authors reach the concept of a cell assembly, where spread activations of neurons function as distributed representations of concepts with which the brain interacts \cite{hebb_organization_1949,palm_cell_1990}. From this, groups like Hofdstater's and Palm's \cite{hofstadter_go_1999,palm_neural_2022} propose these assemblies as the brain's language.

Palm's work goes further and proposes a very complete implementation of an intelligent system working with neural assemblies. The workhorse of his model is associative memory \cite{palm_associative_1980,hecht-nielsen_neurocomputing_1989,palm_memory_1991}. More concretely the Willshaw associative memory model \cite{willshaw_non-holographic_1969}. We will get into the details in the next section, but for now it suffices to state that this very simple binary memory is incredibly efficient both in terms of its information storage capacity, and its implementation in time and energy. This simple component gives rise to a very complete model of cognition, which can solve intelligent tasks in a biologically-constrained manner, and leads to predictions in neuroscience (more on that in the next section).

Despite its tremendous capabilities, the associative memory model has a significant limitation in that it is only able to achieve optimal performance when working with logarithmic sparse data, and real data is nowhere near that sparse (see section \ref{sec:back}). So, research around the model is mostly either theoretical or uses randomly generated sparse data. Thus, the problem of generating suitable representations has been delegated to hypothetical encoders that would play the role of the sensory-specific areas in the cortex. This type of abstraction is common in brain models seen, for example, in Hawkins's notorious model of Hierarchical Temporal Memory \cite{cui_continuous_2016,ahmad_how_2019}.

An encoder that could produce these incredibly sparse representations would unlock the full potential of the associative memory, and by doing so, bring models like Palm's to the aforementioned intersection between understanding the brain and building systems that work well on real data.

So, can it be built? Not easily. To understand why that is so, let us think about the encoding as the vector of activations of a group of neurons. It is not enough for each input to elicit a small number of active neurons. Although such vector would indeed be sparse in the sense that it has a small amount of 1s, if the set of possible inputs always elicits the same subset of neurons, then only a small region of the space is being used, and the zeros are an artificial padding.

So, what we need is for each neuron to have approximately the same, very low probability of firing. Only this will create a sparse but well distributed usage of the space. Additionally, similarity needs to be preserved in the encoding. That is, if two inputs are similar in some sense, the overlap of bits between their codes should be large \cite{Hawkins-et-al-2016-Book,palm_neural_2013}.

All these constraints make the problem just as complex as it is important. And even though some techniques exist to generate sparse codes, as we will see, they are inapt for the logarithmic sparseness constraints imposed by the associative memory. In this paper, we will tackle this problem, and propose a new way to build sparse codes that meets these demanding requirements.

\section{Background in Associative Memory and Neural Assemblies}
\label{sec:back}
Though not as popular in current research, associative memories were once the main focus of neural computation \cite{palm_associative_1980,hecht-nielsen_neurocomputing_1989}. They were mostly set aside because unlike more complex networks such as multilayered perceptrons or convolutional networks they could not perform well on real, challenging tasks.

The idea behind associative memories was to extract the most out of a very simple architecture, and so, by design, these networks are very simple. This makes them incredibly efficient and implementable by extremely fast hardware.

The goal of an associative memory is to associate vectors $\mathbf{x} \in \mathbf{R}^N$ with vectors $\mathbf{y} \in \mathbf{R}^M$. More concretely, it is supposed to store a set of $L$ pairs $\{\left( \mathbf{x}^{(l)}, \mathbf{y}^{(l)} \right) \}_{l=1}^L$ such that when presented with a cue  $\mathbf{x}$ it should be able to output its correspondent associate $\mathbf{y}$.

When the two vectors differ, the memory is said to be doing hetero-association, whereas when a pattern is associated with itself it is called auto-association. Why would a pair be associated with itself? By doing so, a memory can recall a full pattern from a cue vector containing a small portion of it. As we will see later, this use case can become incredibly useful.

Many associative memories have been proposed over the years \cite{palm_associative_1980,hecht-nielsen_neurocomputing_1989}. In general, all fit into one of two groups - feedforward, or recurrent. The former is a simpler machine and computes its output at once. A good example of this type of memory is Willshaw's associative memory \cite{willshaw_non-holographic_1969}. On the other hand, the latter is a dynamical system that evolves until it has converged to a stable state, and a typical example is the Hopfield Network.

Although it may seem counter-intuitive at first, we will see that, under certain conditions, Willshaw's associative memory far outperforms most recurrent memories in terms of information capacity \cite{hecht-nielsen_neurocomputing_1989}. Additionally, its simplicity makes it incredibly efficient both in terms of computation and energy expenditure, which combined with other characteristics make it a model really worth exploring. Given that it is the central memory of this work, from this point onward we will just call it associative memory, and every time we refer to another one we will do so explicitly.

\subsection{The associative memory network}
The associative memory is a feedforward network with binary weights. Given an input vector $\mathbf{x} = \left[ x_1, x_2, \cdots, x_N\right]^T$, its McCulloch  \& Pitts neurons \cite{mcculloch_logical_1943} compute their outputs according to equation \ref{eq:neuron}, where $H$ is the left continuous zero-one step function, and $T_j$ is the neuron-specific threshold.

\begin{equation}
    y_j = H\left( \sum_{i=1}^N w_{ij} x_i - T_j \right) = \begin{cases}
  1 & \sum_{i=1}^N w_{ij} x_i \geq T_j \\
  0 & \text{otherwise}
\end{cases}
    \label{eq:neuron}
\end{equation}

As a learning mechanism, the memory employs typical hebbian learning in that a synapse between two neurons $x_i$ and $y_j$ is present if, during training, the two neurons fire simultaneously \cite{hebb_organization_1949}. More specifically, given a training pair $\left(\mathbf{x},\mathbf{y}\right)$ a general weight $w_{ij}$ will be updated according to equation \ref{eq:hebb}.

\begin{equation}
    w_{ij}^{new} = \begin{cases}
  1 & w_{ij}^{old} = 1\\
  1 & x_iy_j = 1\\
  0 & \text{otherwise}
\end{cases}
    \label{eq:hebb}
\end{equation}

This simple model was shown to have a tremendous storage capacity. In here we will give a brief summary of the storage capacity derivation, with the intent to bring to the reader's attention the key assumptions that make the model work optimally.

Without loss of generality, let us focus on a case where the vectors to be associated are different but have the same size $N$.
Every time we present the memory with a pair of vectors $\left( \mathbf{x}, \mathbf{y} \right)$ we are really storing $\mathbf{y}$, given that the memory requires $\mathbf{x}$ to be presented as a cue. So, in effect the memory is storing $L$ vectors.
Assuming all vectors have exactly $M$ active bits, the information content of a vector equals $\log_2 {N \choose M}$. Therefore, the amount of information stored in the memory will equal $I = L \log_2 {N \choose M}$.

Now, let us assume that the $M$ bits can be in any vector position with the same probability, that is, each dimension is used with a relative frequency of approximately $\frac{M}{N}$. With that, a general weight $w_{ij}$ will equal zero if and only if there is no single pair where $x_i=1$ and $y_j=1$. The probability that $x_i=1$ is $\frac{M}{N}$, just like the probability of $y_i=1$. So, assuming independence, the probability of the synapse being on equals the joint $\frac{M^2}{N^2}$. Therefore, the probability of the weight being off for this particular example is $1 - \frac{M^2}{N^2}$. Now, for the weight to be off after the whole learning, if we assume the samples are independent and identically distributed, then $P \left[w_{ij} = 0 \right] = \left( 1 - \frac{M^2}{N^2} \right) ^ L$. Consequently, $p = P \left[w_{ij} = 1 \right] = 1 - \left( 1 - \frac{M^2}{N^2} \right) ^ L$.

At recall time, the memory will produce a spurious bit $y_k=1$ in the output if and only if $\forall_{i=1}^N x_i = 1 : w_{ik} = 1$. The probability of this event is given by $p^M$.

Setting the limit for the number of acceptable spurious bits to one (i.e. assuming $(N-M)p^M = 1$), it is possible to get concrete numbers for the most important quantities. For example, the optimal amount of active bits per pattern is $M = \log_2 N - 2$. This tells us that the vectors must be logarithmic sparse. Additionally, the maximum amount of patterns that can be stored is given by $L = \log_2 \left( \frac{N^2}{M^2} \right)$. Finally, the information capacity under these optimal assumptions is $I = N^2 \log_2 2$, which for an $N \times N$ bit matrix yields a per bit capacity of $\log_2 \approx 69.31\%$. This value is much higher than most associative memories, and it is especially significant given the simplicity of the model, and its biological plausibility. All in all, the model only uses well-established hebbian learning, it can be implemented in neural hardware, and, like the brain, it has tremendous energy efficiency \cite{palm_neural_2022}.

\subsection{Neural assemblies via associative memory}

Neurons that extract particular features have been found in many studies across the brain \cite{palm_neural_2022}. Yet, no studies find neurons that solely encode a particular concept. Additionally, if there were such neurons, the brain would not be very robust given that neurons die everyday, so us humans could be forgetting concepts routinely. In contrast, a much more plausible, robust, and efficient way to encode information is via distributed representations \cite{Hawkins-et-al-2016-Book,ahmad_how_2019}.

Adding this fact, to the nowadays widely accepted presence of hebbian learning in the brain \cite{citri_synaptic_2008,mi_synaptic_2017,fiebig_spiking_2017}, many authors reach the concept of a neural assembly \cite{palm_cell_1990,gerstein_neuronal_1989,singer_neuronal_1997,wallace_chasing_2010,palm_cell_2014}. This concept corresponds to a set of neurons that have become so wired together through learning that activation of a subset of them leads to the activation of the whole group.

Being distributed representations of concepts, these assemblies are constituted by single neurons that can participate in many other assemblies.

This seemingly simple concept can act as the bridge between between the experimental neuroscientific level, and the information processing level \cite{hofstadter_go_1999,palm_neural_2022}. Coarser than single neurons, assemblies lend themselves better to higher-level reasoning about cognitive and thought processes.

As it is based on hebbian learning and distributed representations, associative memory is a great candidate  to model cell assemblies, and Palm's work has focused on that \cite{palm_cell_1990,palm_cell_2014}. Concretely, the ignition of an assembly can viewed as an instance of auto-association, whereas sequencing of these assemblies corresponds to hetero-association. Adding the biologically plausible mechanism of threshold control \cite{palm_cell_2014}, Palm's group has built models with high explanatory capability that view the cortex as a group of associative memories. This type of idea has been applied for instance to language understanding \cite{hutchison_combining_2005,knoblauch_associative_2005}. And, by the nature of the associative memory model, it has led to widely confirmed neuroscientific predictions. Among which we highlight two: first, that many brain areas must be involved in most tasks \cite{arieli_dynamics_1996,markram_blue_2006}; second, as we have seen in the previous subsection, that the distributed representations in the brain should be sparse \cite{olshausen_emergence_1996,willmore_characterizing_2001,waydo_sparse_2006,quiroga_invariant_2005,quiroga_sparse_2008,quiroga_concept_2012}.

The confirmed prediction of sparse representations generates more confidence in the associative memory as a fundamental building block of the cortex. However, as we have stated before, this sparse coding requirement \cite{palm_computing_1987,palm_associative_1994,palm_neural_2013,sa-couto_storing_2020} is also its main curse for practical applicability. In the next section, we will attempt to solve this problem.

\section{Proposed Network}
A group of neurons can represent vectors $\mathbf{x} \in \mathbb{R}^N$ by learning a set of prototypes $\{ \mathbf{w}_1, \mathbf{w}_2, \cdots, \mathbf{w}_N \}$ that best represent them. Having learned an alphabet, an encoding is a binary presence vector $\mathbf{z}$ that indicates to which prototype the current input is most similar to, according to some distance function $D \left( \mathbf{w}_i, \mathbf{x} \right)$. Concretely, each element of the encoding vector $z_i$ is the indicator function of equation \ref{eq:wta}, and this type of neuron group is usually called a winner-takes-all competitive layer \cite{hecht-nielsen_neurocomputing_1989,sa-couto_attention_2019}.

\begin{equation}
z_i = \mathbb{I} \left( i = \argmin_{1 \leq j \leq N} D \left( \mathbf{w}_j, \mathbf{x} \right)  \right)
\label{eq:wta}
\end{equation}

Competitive layers have been analyzed extensively and their learning can be implemented in a biologically plausible manner through lateral inhibition \cite{hecht-nielsen_neurocomputing_1989}. Abstracting away the details, the process is quite simple. Given an input, the neurons compete to represent it, and the one that best fits in a nearest neighbor sense is the winner, and gets to update. All the others stay the same. The update rule is the simple iterative mean in equation \ref{eq:comp_learn}, and it makes the winning neuron even more suited to represent the current input.

\begin{equation}
    \mathbf{w}_i^{new} = \mathbf{w}_i^{old} + \alpha \left( \mathbf{x} - \mathbf{w}_i^{old} \right) z_i
    \label{eq:comp_learn}
\end{equation}

With a ``winner-takes-all'' approach the encoding is a one-hot vector indicating the winner. Therefore, these encodings are sparse in the sense that there is only one active bit and lots of zeros. Yet, nothing guarantees that the neurons will fire equally frequently. Specifically, when sampling an input vector $\mathbf{x}$, we would like, to have weight vectors $\{ \mathbf{w}_1, \mathbf{w}_2, \cdots, \mathbf{w}_N \}$, such that every neuron wins the competition with probability $\frac{1}{N}$. Fortunately, this problem of equiprobability was solved by Desieno \cite{desieno_adding_1988}, by penalizing neurons that fire too often.

This activity-dependent bias is also biologically plausible, and can be implemented with a local learning rule. Specifically, we use equation \ref{eq:bias_1} to keep an estimate of how active each neuron has been. This average activity $f_i$ is then used to compute the bias $b_i$ in equation \ref{eq:bias_2} (where $\gamma$ controls the weight of the bias in the competition). Having the bias, the winning neuron is not the one that minimizes the distance $D$, but the one that minimizes $D \left( \mathbf{w}_i, \mathbf{x} \right) - b_i$.

\begin{equation}
    f_i^{new} = f_i^{old} + \beta \left( z_i - f_i^{old} \right)
    \label{eq:bias_1}
\end{equation}

\begin{equation}
    b_i = \gamma \left( \frac{1}{N} - f_i \right)
    \label{eq:bias_2}
\end{equation}

Adding this bias, the winner-takes-all competitive layer can generate encodings with neurons that fire sparsely and with very similar frequency. So, have we reached the goal? Not exactly, since the key property of similarity preservation is missing. The one-hot encodings are quasi-symbolic representations of their inputs in the sense that there is no partial similarity between them. Concretely, when there is a very small distance between two inputs, they will lead to the same winner, and therefore to the same encoding. In this case, there will be an overlap leading to a similarity score of $1$ bit. 

However, when two inputs differ to a point there will never be an overlap, so similarity will always be zero regardless of how similar they may be. With that in mind, there can be severe limitations on the amount of prototypes to extract, and, as a consequence, on the richness of the domain we can model. When too many prototypes exist, there will inevitably be some similar ones. In this case, there will be a high probability of two very similar inputs resulting in two different encodings.

At this point, what we already have is a way to learn groups of neurons that encode some input into sparse and equal frequency firing codes. Yet, we do not have a way to allow for similarity preservation in them. Now, this method will generate codes with a single bit. If, instead of using one group of neurons to encode an input, we use $M$, then our vector will be just as sparse in terms of percentage of active bits, and it will also be just as uniformly used. We can leverage this principle to add active bits to our code, and thus allow for a richer description. 

However, if the input to all the neuron groups is the same, then, their learning will be nearly identical, and the use of $M$ bits instead of one will be redundant. To solve this problem we borrow a principle from biology and make each neuron group focus on a randomly assigned, localized receptive field. By doing so, we will cause each group to model an area of the visual field, thus breaking the problem down into less complex ones. In the final code, the winner of each group will represent the content of a visual region, thus working as a binary indicator of a learned feature.

\begin{figure}
  \centering
  \includegraphics[width=0.7\textwidth]{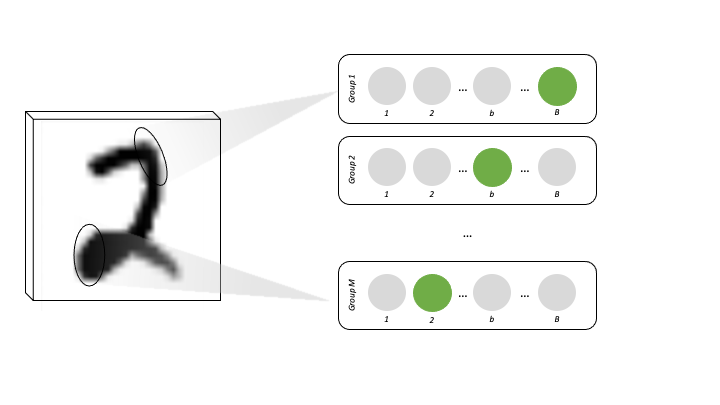}
  \caption{An image is encoded by a set of $M$ neuron groups. Each group focuses on a receptive field, and tries to represent its contents. To that end, inside each group neurons compete to represent the current input. The winner takes it all and fires alone. In the end, the vector of all activations will be a binary sparse code, where each neuron fires approximately equally often.}
  \label{fig:fig0}
\end{figure}

Figure \ref{fig:fig0} presents a schematic view of the proposed encoder network. In this example, we depict three of $M$ neuron groups, each containing $B$ elements. To achieve an $H$-dimensional encoder, we can, for instance, choose $B$ such that it approximately equals $\frac{H}{M}$. Also depicted in the figure are the receptive fields of the first and last groups. Each focuses on an area of the image and uses its winning neuron to encode the content in it. With this arrangement, the sparse codes will be able to represent more levels of similarity between different inputs, and thus our problem is addressed. 

With the fully developed mathematical details of our proposed encoder, in the next section we turn to an empirical evaluation. We will try to see whether or not our intuitions hold, how the proposed model compare to equivalent approaches, and finally, if it makes the use of associative memory with real data possible.

\section{Experimental Analysis}
\begin{figure}
  \centering
  \includegraphics[width=0.4\textwidth]{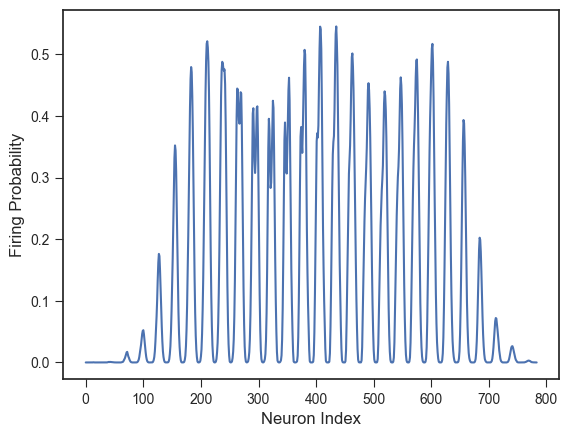}
  \includegraphics[width=0.4\textwidth]{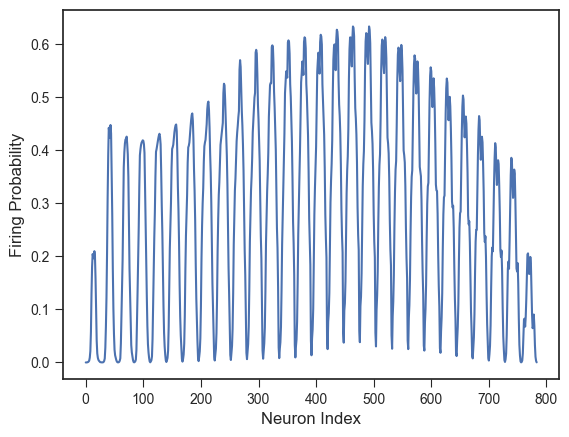}
  \caption{Assuming each neuron encodes one pixel, we can plot their usage for regular images. By doing so, we can see that with real data, the firing probability is neither small, nor similar across neurons. On the left we can observe that on the MNIST dataset, whereas on the right we present the same for Fashion-MNIST.}
  \label{fig:fig1}
\end{figure}

Throughout this section we will explore the capabilities of our proposed solution using two well-known visual pattern recognition sets: MNIST \cite{lecun_mnist_nodate} and Fashion-MNIST \cite{xiao2017/online}. Both are composed of $70000$ grayscale images of $28 \times 28$ pixels. Assuming one neuron per pixel, we took these examples, and plotted the relative firing frequencies for both data sets in Figure \ref{fig:fig1}. Analyzing the results we see exactly why we said that real data is neither sparse nor well distributed across vector dimensions.

\subsection{Comparison to a Deep Learning baseline}
To properly evaluate the quality of the proposed codes for the associative memory we need a baseline. To that end we chose deep autoencoders \cite{goodfellow_deep_2016} that not only minimize a reconstruction loss, but also minimize the average firing frequency penalty on the form of the batch-wise KL-divergence that is presented in equation \ref{eq:KL} (where $\rho$ is the desired firing relative frequency, $L$ is the batch size, and $H$ is the size of the encoding layer). Additionally, to ensure a binary encoding, we use a Gumbel Softmax activation function for the middle hidden layer \cite{jang_categorical_2016}.

\begin{equation}
    \sum_{l=1}^{L}  \sum_{i=1}^{H}  \left( \rho * log \left(\frac{\rho}{h_{li}} \right) + \left(1-\rho \right) * \log \left(  \frac{1-\rho}{1-h_{li}} \right) \right)
\label{eq:KL}
\end{equation}

Having established the baseline architecture, there are many hyper-parameters that can be tuned. The size of the encoding layer $H$ is chosen at random to be one of three options: 512, 1024, and 2048. From there we can compute the desired number of active neurons to be approximately $M=\log_2 \left( \frac{H}{4} \right)$ (see section \ref{sec:back}). Having $M$, we can define the target firing relative frequency $\rho = \frac{
M}{H}$. The number of layers is a random even number between two and eight, and the number of neurons in each layer changes linearly from input size to $H$. For training we also sample different: numbers of epochs, learning rates, reconstruction losses, batch sizes, and relative strengths of the sparsity constraint\footnote{The code for all simulations can be obtained upon email request to the first author}.

\begin{figure}
  \centering
  \includegraphics[width=0.7\textwidth]{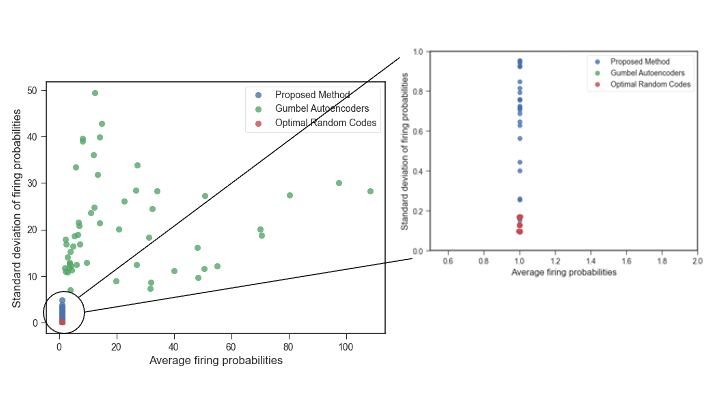}
  \caption{The mean and standard deviation of neurons' firing probabilities, on the MNIST dataset, for several runs of three models: Gumbel autoencoders baseline, the proposed method, and optimal random codes.}
  \label{fig:fig2}
\end{figure}

Each green dot in figure \ref{fig:fig2} corresponds to a fully trained autoencoder after going through the aforementioned sampling steps. The blue dots represent runs of our proposed method where all the hyper-parameters are also randomly sampled from an interval of reasonable values. Finally, the red dots correspond to randomly generated data that perfectly fits the requirements of associative memory. Analyzing the plot, we can see that our approach not only clearly beats the autoencoder baseline, but also that it gets extremely close to the optimal ceiling.

\begin{figure}
  \centering
  \includegraphics[width=0.34\textwidth]{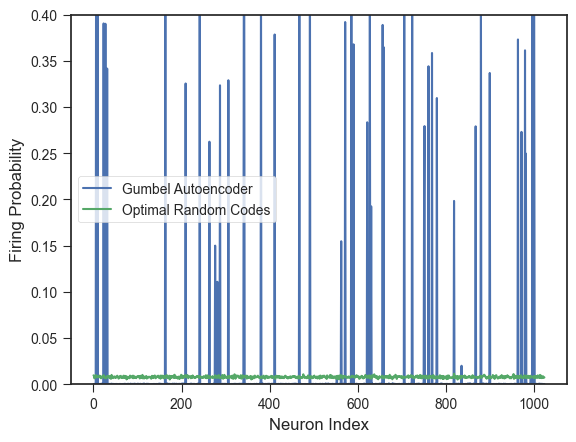}
  \includegraphics[width=0.34\textwidth]{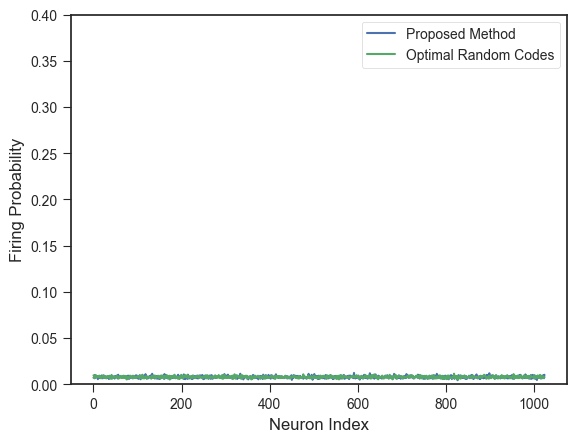}
  \includegraphics[width=0.34\textwidth]{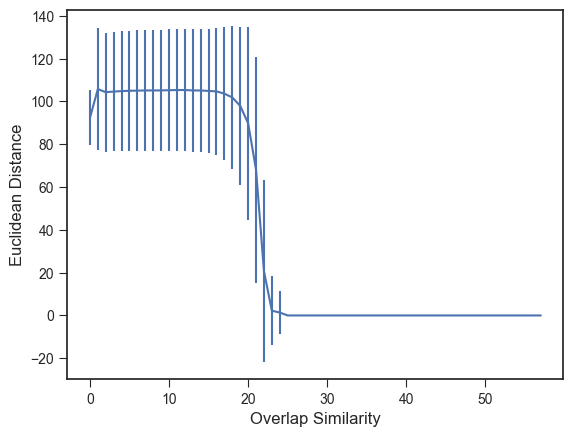}
  \includegraphics[width=0.34\textwidth]{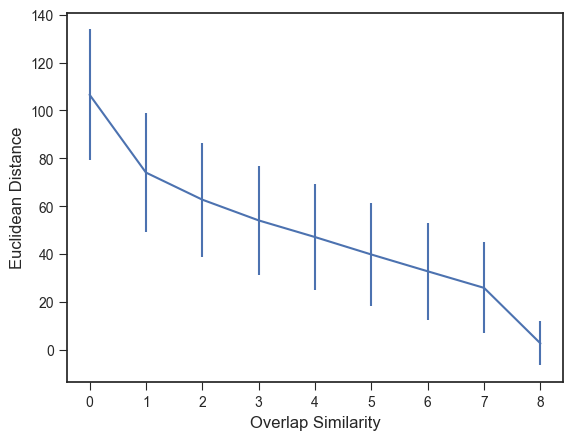}
  \caption{The top row presents the individual neurons' firing probabilities for the best autoencoder (left) and the best run of the proposed solution (right). Additionally, both are compared to the probabilities of optimal random codes. The proposed method is much closer to the goal, the lines almost fully overlap. On the bottom row we present the relation between euclidean distance in the images' space and the bit overlap between the codes. Though both preserve some similarity, the proposed method (on the right) has a gradual decrease in distance as similarity increases, whereas autoencoders (on the left) exhibit almost a binary behavior.}
  \label{fig:fig3}
\end{figure}

If we take the best runs from each of the three types of codes (ensuring that they all have the same dimensionality), we can make the more detailed comparison that is presented in figure \ref{fig:fig3}. By looking at each neuron's firing frequency, we see just how close the proposed solution is to being optimal. Furthermore, we can see that it better preserves the similarity between images in the encoded space.

\subsection{Comparison to classical Sparse Coding}
Another very common approach to sparse coding is that of dictionary learning \cite{mairal_online_2009}. In this setting, we learn a dictionary of features $\mathbf{V}$, and learn the best possible sparse encoding $\mathbf{U}$, such that $\mathbf{U}\mathbf{V}$ approximates the inputs $\mathbf{X}$. This is translated into the loss function written in equation \ref{eq:sparse_coding}. Both $\mathbf{U}$ and $\mathbf{V}$ must be found through optimization, and in general, alternated coordinate descent is used.

\begin{equation}
    \frac{1}{2} \| \mathbf{X} - \mathbf{U}\mathbf{V}\|_F^2 + \alpha \|\mathbf{U}\|_1
    \label{eq:sparse_coding}
\end{equation}

By doing a random search through the space of hyper-parameters we found an apt version of this classical sparse coding technique which will be used as a term of comparison to our approach. To level the field, we made sure all codes have the same dimensionality of $1024$ neurons.

\subsubsection{Neurons' firing frequencies}
\begin{figure}
  \centering
  \includegraphics[width=0.4\textwidth]{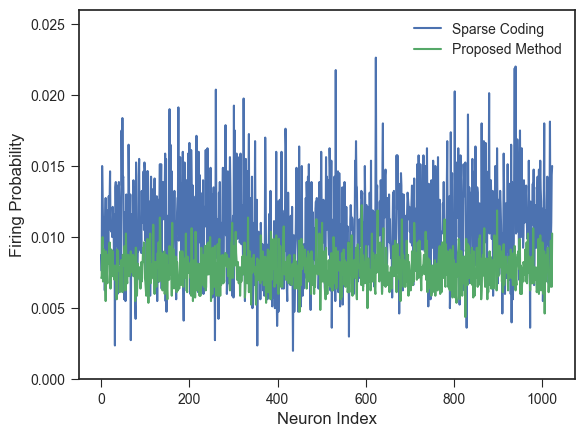}
  \includegraphics[width=0.4\textwidth]{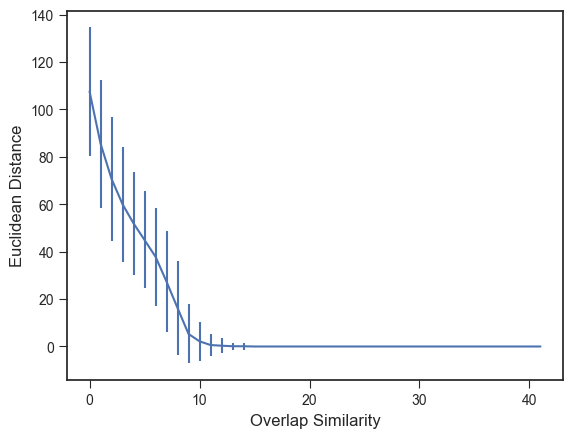}
  \caption{On the left we compare the individual neuron's firing probabilities between the proposed approach and coordinate-descent-based sparse coding. On the right, we plot the relation between input space distance and overlap similarity for classical sparse coding.}
  \label{fig:fig4}
\end{figure}

Once again, we compare the neuron usage for both codes. Analyzing figure \ref{fig:fig4}, we see that the difference between the two approaches is much smaller than the difference between our approach and the autoencoder baseline. Furthermore, we see that classical sparse coding is also able to preserve similarity. However, as we will see next, these seemingly small differences carry a tremendous cost when dealing with the associative memory.

\subsubsection{Storing in auto-association}
\begin{figure}
  \centering
  \includegraphics[width=0.34\textwidth]{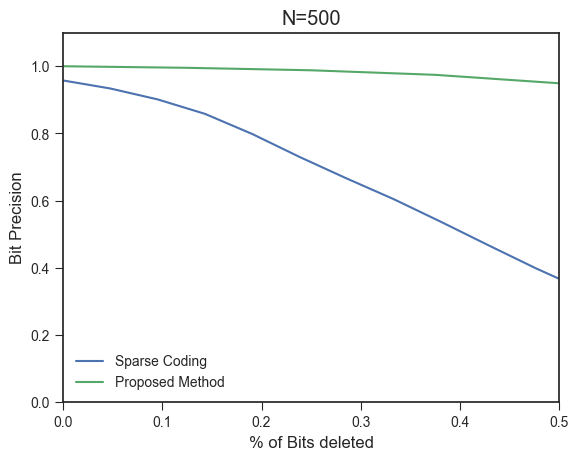}
  \includegraphics[width=0.34\textwidth]{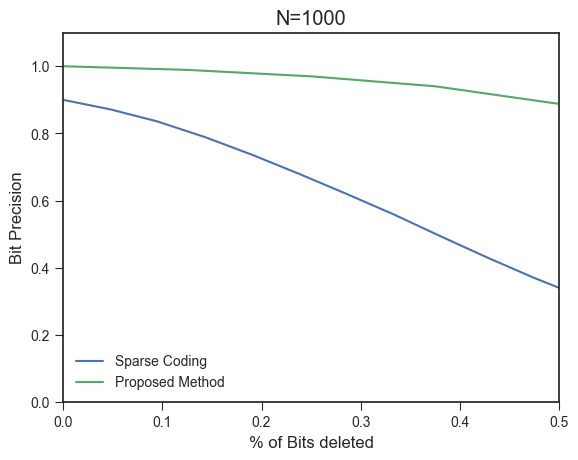}
  \includegraphics[width=0.34\textwidth]{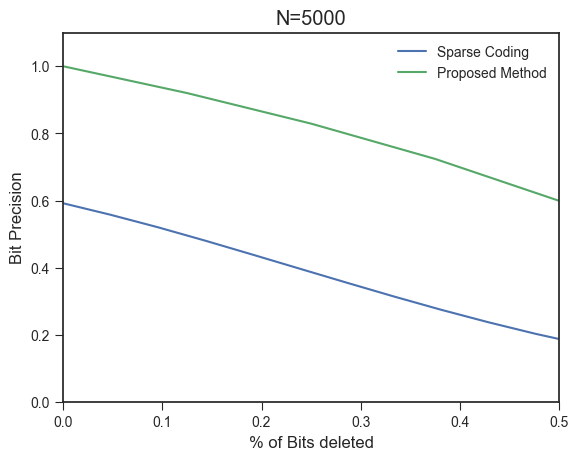}
  \includegraphics[width=0.34\textwidth]{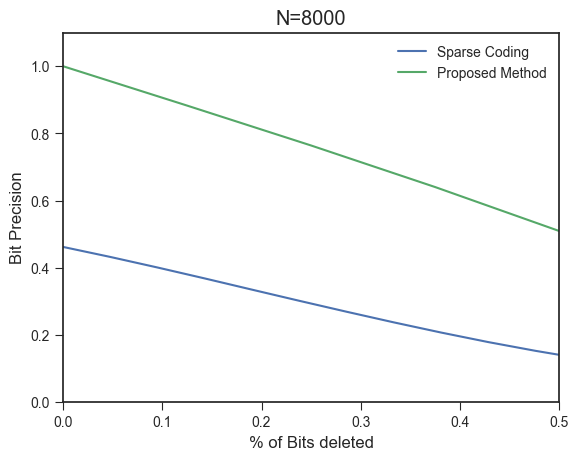}
  \caption{As the number of stored examples increases, the proposed codes maintain a high bit precision whereas classical sparse codes lead to rapidly decreasing precision.}
  \label{fig:fig6}
\end{figure}

Now that we are sure that both techniques approximately fit the memory's requirements we can take the two sets of encodings and store them in a memory. We start with auto-association where each pattern is associated with itself, and then we progressively delete bits and see whether the memory is able to complete the missing information. At a first glance, two measures seem important: bit recall, and bit precision. However, after a more detailed analysis, we found out that the memory never loses bits that are presented as cues, so recall is always at $100\%$. For that reason, in figure \ref{fig:fig6}, we only present the average bit precision for both techniques, across $30$ runs, as the number of deleted bits increases. The results clearly favor the proposed method, regardless of how full the memory is. Furthermore, precision is still high even when the memory is filled with a number of patterns that is approximately $8$ times its number of neurons (in this case $1024$).

\begin{figure}
  \centering
  \includegraphics[width=0.5\textwidth]{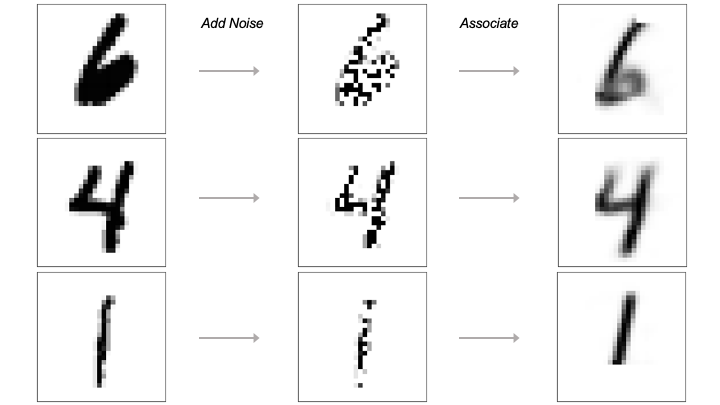}
  \caption{A few examples of using the auto-association memory with our codes to do pattern completion.}
  \label{fig:fig7}
\end{figure}

A typical use case of auto-association is that of pattern completion, or interpolation. For illustrative purposes we provide figure \ref{fig:fig7} where each pixel has a $50\%$ chance of being turned to zero. Then, we encode the noisy patterns, ask the memory to auto-associate them, and then we reconstruct an image from the code. The results are very positive, and, in general, bit precision is a harsher performance measure than visual inspection of the reconstruction quality.

\subsubsection{Storing in hetero-association}
\begin{figure}
  \centering
  \includegraphics[width=0.8\textwidth]{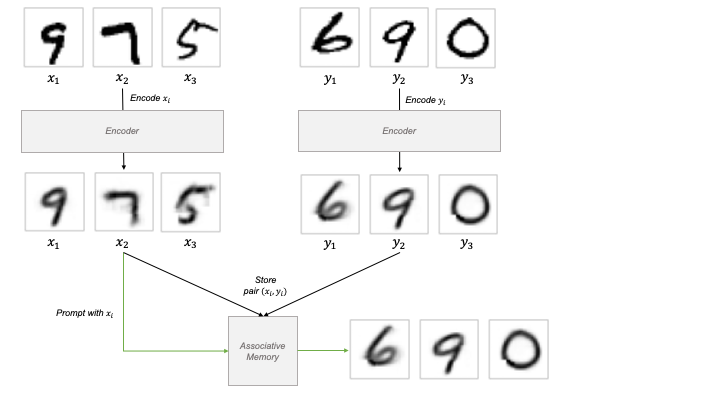}
  \caption{Each pair of images to be associated is first encoded, and then presented to the memory. Having a trained memory, we can present it with a prompt image, and it will try to reconstruct its pairing. The figure presents some real experimental examples.}
  \label{fig:fig9}
\end{figure}

All the experiments with auto-association are important, however the most difficult task the associative memory can do is hetero-association. Experiments with the classical sparse coding technique showed it to be completely inadequate for this purpose. So, we decided to compare our approach with optimal random codes. Figure \ref{fig:fig9} illustrates the hetero-associative pipeline. To store a pair of images in memory we encode them, and then present the respective sparse codes to the memory in the form of associate pairs. Then, we measure hetero-associative quality by prompting the memory with one of the pair's members to see whether or not the memory is able to recall the correct associate. The examples depicted therein are real, and were achieved with our proposed approach.

\begin{figure}
  \centering
  \includegraphics[width=0.4\textwidth]{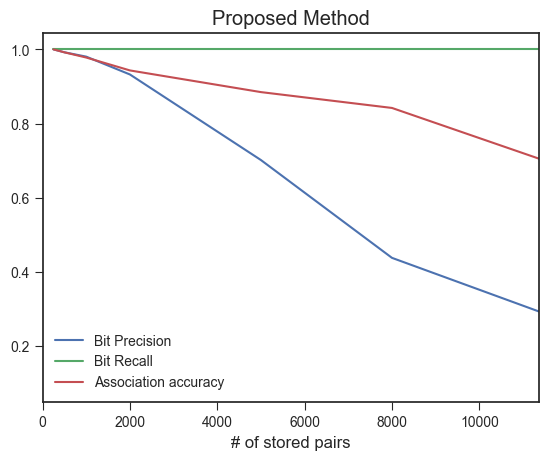}
  \includegraphics[width=0.4\textwidth]{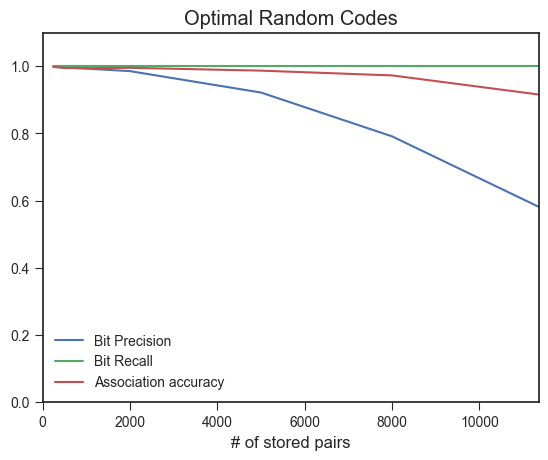}
  \caption{On the left we track bit recall, bit precision, and association accuracy for a hetero-associative memory trained with the proposed codes. On the right we track the same measures for optimal random codes. Though not optimal, the proposed codes lead the memory to a very competitive behavior.}
  \label{fig:fig8}
\end{figure}

Unlike in auto-association, bit recall is not guaranteed to be perfect as we are not associating a pattern with itself. Also, it becomes interesting to measure the association accuracy of the memory, and find out how many times it recalls a completely wrong pattern. We define this measure by finding the code which most overlaps with the memory's output. If it is the wrong code, then we count an error. In figure \ref{fig:fig8}, we depict the three key measures as the number of stored patterns increases. Our approach holds up extremely well when compared to the optimal random codes, especially when it comes to bit recall, and association accuracy. The fact that usage is not perfectly uniform leads to spurious correlation that cause some wrong neurons to fire when the memory is extremely full.

\subsection{Extending to a more complex data set}
To make sure our analysis is valid and generalizable, we decided to conduct very similar experiments on a more complex data set - Fashion-MNIST.

\begin{figure}
  \centering
  \includegraphics[width=0.7\textwidth]{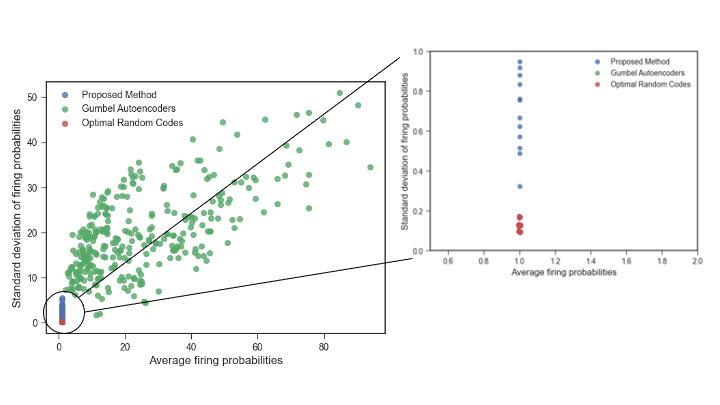}
  \caption{The mean and standard deviation of neurons' firing probabilities, on the Fashion-MNIST dataset, for several runs of three models: Gumbel autoencoders baseline, the proposed method, and optimal random codes.}
  \label{fig:fig10}
\end{figure}

\begin{figure}
  \centering
  \includegraphics[width=0.34\textwidth]{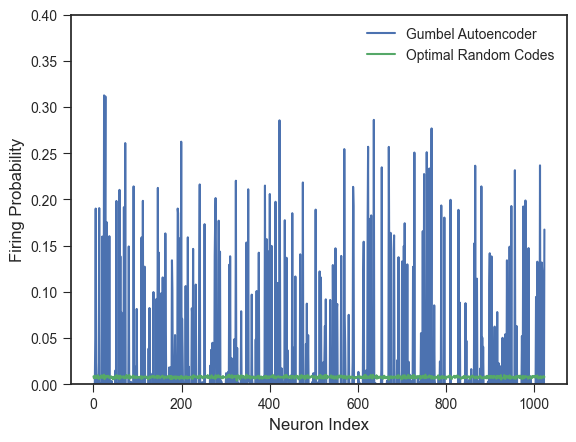}
  \includegraphics[width=0.34\textwidth]{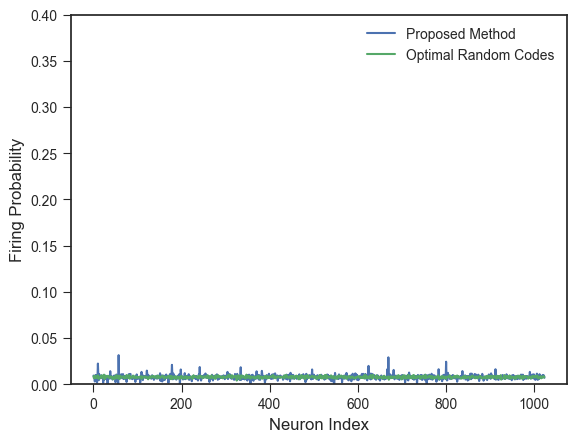}
  \includegraphics[width=0.34\textwidth]{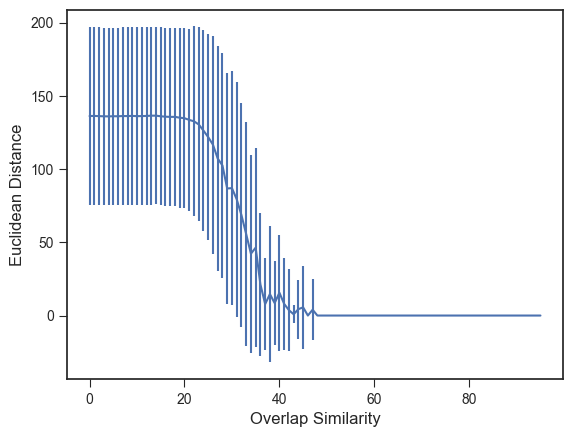}
  \includegraphics[width=0.34\textwidth]{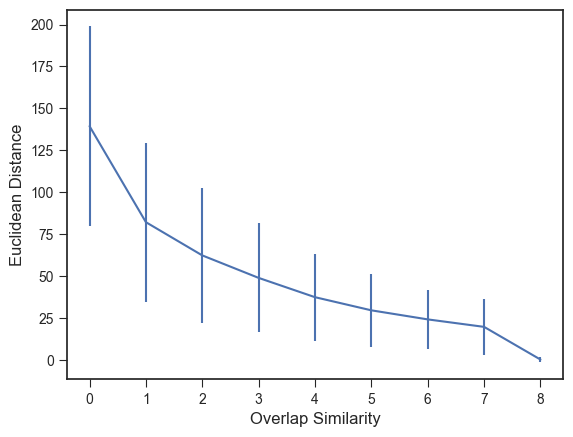}
  \caption{A replica of figure \ref{fig:fig3} but for the Fashion-MNIST data set. The same conclusions hold: the proposed approach presents a clear improvement over the baseline.}
  \label{fig:fig11}
\end{figure}

Figures \ref{fig:fig10} and \ref{fig:fig11} present the comparison to the autoencoder baseline both in terms of neuronal relative firing frequency and similarity preservation. The conclusions are exactly the same as for MNIST as the figures look almost like replicas of figures \ref{fig:fig2} and \ref{fig:fig3}.

\begin{figure}
  \centering
  \includegraphics[width=0.42\textwidth]{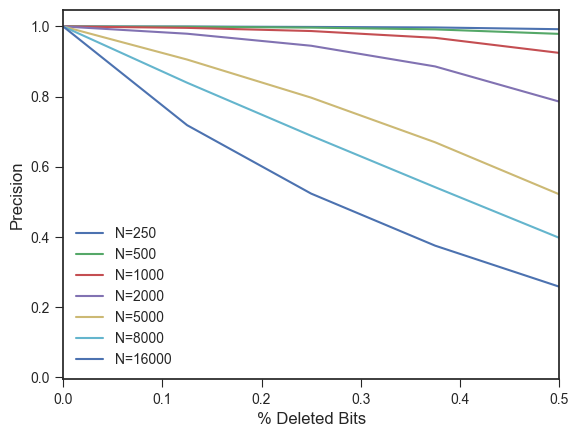}
  \caption{The proposed coding manages to keep a high bit precision even though the number of stored patterns largely supersedes the number of units ($1024$ in this case).}
  \label{fig:fig12}
\end{figure}

Figure \ref{fig:fig12} shows how the memory's bit precision in auto-association is very high even for extremely harsh conditions where it is incredibly full, and a large percentage of cue bits has been deleted. 

\begin{figure}
  \centering
  \includegraphics[width=0.4\textwidth]{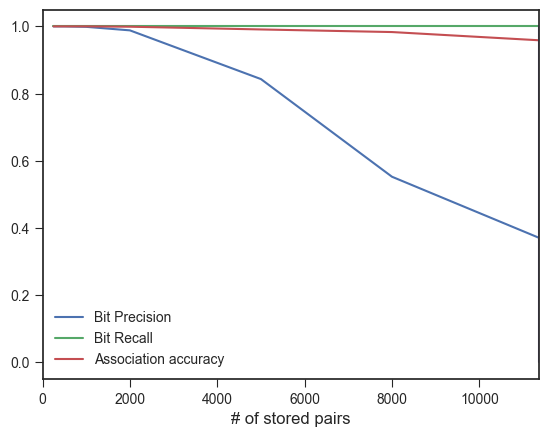}
  \includegraphics[width=0.4\textwidth]{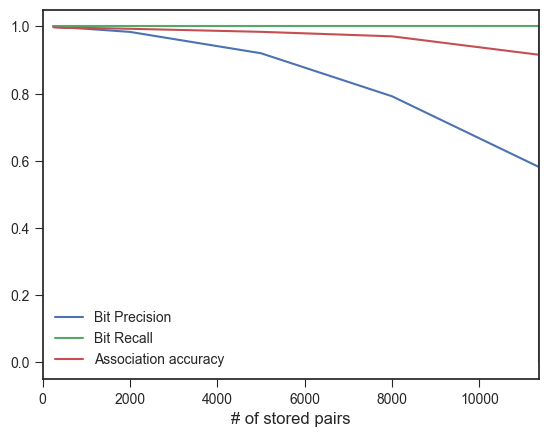}
  \caption{Once again, we use three key performance measures to compare the associative memory performance using the proposed codes and optimal random codes. Yet, this time we do it on the Fashion-MNIST data set.}
  \label{fig:fig13}
\end{figure}

Finally, figure \ref{fig:fig13} provides the same hetero-associative analysis we did for MNIST (see figure \ref{fig:fig8}). Once again, our approach is capable not only of mostly correct associations, but the results are even more competitive in terms of association accuracy. This phenomenon is most likely due to the fact that the data set is more varied when compared to the more uniform MNIST.

\section{Conclusion}

Even though artificial intelligence is getting better, no system achieves the flexibility and adaptability of the human brain. Therefore, computational neuroscience is a key research field that not only can help discover the mysteries behind it, but also serve as inspiration for new and better intelligent systems. However, most biologically inspired systems have trouble competing with techniques like Deep Learning on real data. Only by bridging this gap can we take the field to the next level.

The concept of neural cell assemblies, first advanced by Hebb, has led to several brain models with lots of potential. For instance, Palm's work implements this concept through binary associative memory. This seminal model makes neuroscientific predictions, and shows tremendous explanatory power for the mechanics of thought processes, and cognitive capabilities.

Yet, there is a catch. Associative memory only performs well when dealing with logarithmic sparse, equal frequency, similarity-preserving codes. Given that real data is far from having these properties, and that no encoders exist that can adequately map it in such manner, most work around it was done with randomly generated data.

We set out with the intent to build an encoder that could take real visual data, and represent it as sparse codes that met all the memory's tough constraints. Namely that all the dimensions in the code should be used with approximately equal probability, and similarities in the input space should be preserved in the code space.

We defined a group of neuron to be a ``winner-takes-all'' competitive layer, where equal probability of firing is ensured by Desieno's idea of activity-dependent bias. With that, we noted that, by looking at active/inactive neurons as bits, a group of neurons would result in a sparse, equal-frequency code.

Then, we observed that to represent a complex domain, we would require a very large group. As it gets larger, features get more specific, but the one-hot encodings cannot represent the partial similarity. To surpass this issue, we borrowed the concept of receptive field from biology, and made several groups of neurons with each specializing on a particular region of the visual field. Putting all together, we got a biologically plausible distributed network that encodes images into sparse codes that are suitable for associative memory.

To test our intuitions, we took the two well-known data sets MNIST and Fashion-MNIST, and compared the codes produced by our method with two alternative approaches. Both coordinate descent sparse coding and Gumbel Softmax autoencoders exhibited worse properties of sparsity, uniformity of firing probability, and similarity preservation when compared to the proposed method. Additionally, the proposed method came extremely close to optimal random codes. Having the codes, we moved on to testing them with the associative memory. Both in auto- and hetero-association, not only can we say that the proposed method beat the baselines, but also that it approached optimality.

In summary, by achieving the proposed goals, our approach opened the doors for further exploration of brain models that use cell assemblies and associative memory. Future work may open this avenue or further develop the encoder in one of two ways. First, we may compose groups of neurons to encode even richer features, and generalize to even more complex tasks. Second, the encoder can be applied to other types of perceptual data like speech without great changes.

\section*{Acknowledgments}
The first author dedicates this work to Margarida, without whom it would never have been possible. Additionally, we would like to acknowledge support for this project from the Portuguese Foundation for Science and Technology (FCT) with a doctoral grant SFRH/BD/144560/2019 awarded to the first author, and the general grant UIDB/50021/2020. The Foundation had no role in study design, data collection and analysis, decision to publish, or preparation of the manuscript. The authors declare no conflicts of interest. Code and data for all the experiments can be obtained by email request to the first author.

\bibliographystyle{unsrt}  
\bibliography{references}

\begin{thebibliography}{10}

\bibitem{zador_toward_2022}
Anthony Zador, Blake Richards, Bence Ölveczky, Sean Escola, Yoshua Bengio,
  Kwabena Boahen, Matthew Botvinick, Dmitri Chklovskii, Anne Churchland,
  Claudia Clopath, James DiCarlo, Surya Ganguli, Jeff Hawkins, Konrad Koerding,
  Alexei Koulakov, Yann LeCun, Timothy Lillicrap, Adam Marblestone, Bruno
  Olshausen, Alexandre Pouget, Cristina Savin, Terrence Sejnowski, Eero
  Simoncelli, Sara Solla, David Sussillo, Andreas~S. Tolias, and Doris Tsao.
\newblock Toward {Next}-{Generation} {Artificial} {Intelligence}: {Catalyzing}
  the {NeuroAI} {Revolution}, October 2022.
\newblock arXiv:2210.08340 [cs, q-bio].

\bibitem{hebb_organization_1949}
D.O. Hebb.
\newblock {\em The {Organization} of {Behavior}}.
\newblock New York, John Wiley \& Sons, 1949.

\bibitem{stuart_active_1994}
Greg~J. Stuart and Bert Sakmann.
\newblock Active propagation of somatic action potentials into neocortical
  pyramidal cell dendrites.
\newblock {\em Nature}, 367(6458):69--72, 1994.
\newblock Publisher: Nature Publishing Group.

\bibitem{citri_synaptic_2008}
Ami Citri and Robert~C. Malenka.
\newblock Synaptic plasticity: multiple forms, functions, and mechanisms.
\newblock {\em Neuropsychopharmacology}, 33(1):18--41, 2008.
\newblock Publisher: Nature Publishing Group.

\bibitem{oreilly_six_1998}
Randall~C. O'Reilly.
\newblock Six principles for biologically based computational models of
  cortical cognition.
\newblock {\em Trends in cognitive sciences}, 2(11):455--462, 1998.
\newblock Publisher: Elsevier.

\bibitem{szatmary_spike-timing_2010}
Botond Szatmáry and Eugene~M. Izhikevich.
\newblock Spike-timing theory of working memory.
\newblock {\em PLoS computational biology}, 6(8):e1000879, 2010.
\newblock Publisher: Public Library of Science San Francisco, USA.

\bibitem{barak_working_2014}
Omri Barak and Misha Tsodyks.
\newblock Working models of working memory.
\newblock {\em Current opinion in neurobiology}, 25:20--24, 2014.
\newblock Publisher: Elsevier.

\bibitem{mi_synaptic_2017}
Yuanyuan Mi, Mikhail Katkov, and Misha Tsodyks.
\newblock Synaptic correlates of working memory capacity.
\newblock {\em Neuron}, 93(2):323--330, 2017.
\newblock Publisher: Elsevier.

\bibitem{fiebig_spiking_2017}
Florian Fiebig and Anders Lansner.
\newblock A spiking working memory model based on {Hebbian} short-term
  potentiation.
\newblock {\em Journal of Neuroscience}, 37(1):83--96, 2017.
\newblock Publisher: Soc Neuroscience.

\bibitem{palm_cell_1990}
Günther Palm.
\newblock Cell assemblies as a guideline for brain research.
\newblock {\em Concepts in Neuroscience}, 1:133--147, 1990.

\bibitem{hofstadter_go_1999}
Douglas~R. Hofstadter.
\newblock {\em Gödel, {Escher}, {Bach}: an eternal golden braid}.
\newblock Basic Books, New York, 20th anniversary ed edition, 1999.

\bibitem{palm_neural_2022}
Günther Palm.
\newblock {\em Neural {Assemblies}: {An} {Alternative} {Approach} to
  {Classical} {Artificial} {Intelligence}}.
\newblock Springer Nature, 2022.

\bibitem{palm_associative_1980}
Günther Palm.
\newblock On associative memory.
\newblock {\em Biological cybernetics}, 36(1):19--31, 1980.
\newblock Publisher: Springer.

\bibitem{hecht-nielsen_neurocomputing_1989}
Robert Hecht-Nielsen.
\newblock {\em Neurocomputing}.
\newblock Addison-Wesley Longman Publishing Co., Inc., 1989.

\bibitem{palm_memory_1991}
G.~Palm.
\newblock Memory capacities of local rules for synaptic modification.
\newblock {\em Journal of Concepts in Neuroscience}, 2(1):97--128, 1991.

\bibitem{willshaw_non-holographic_1969}
David~J. Willshaw, O.~Peter Buneman, and Hugh~Christopher Longuet-Higgins.
\newblock Non-holographic associative memory.
\newblock {\em Nature}, 222(5197):960--962, 1969.
\newblock Publisher: Nature Publishing Group.

\bibitem{cui_continuous_2016}
Yuwei Cui, Subutai Ahmad, and Jeff Hawkins.
\newblock Continuous online sequence learning with an unsupervised neural
  network model.
\newblock {\em Neural Computation}, 28(11):2474--2504, November 2016.
\newblock arXiv:1512.05463 [cs, q-bio].

\bibitem{ahmad_how_2019}
Subutai Ahmad and Luiz Scheinkman.
\newblock How {Can} {We} {Be} {So} {Dense}? {The} {Benefits} of {Using}
  {Highly} {Sparse} {Representations}, April 2019.
\newblock arXiv:1903.11257 [cs, stat].

\bibitem{Hawkins-et-al-2016-Book}
J.~Hawkins, S.~Ahmad, S.~Purdy, and A.~Lavin.
\newblock Biological and machine intelligence (bami).
\newblock Initial online release 0.4, 2016.

\bibitem{palm_neural_2013}
Günther Palm.
\newblock Neural associative memories and sparse coding.
\newblock {\em Neural Networks}, 37:165--171, 2013.
\newblock Publisher: Elsevier.

\bibitem{mcculloch_logical_1943}
Warren~S. McCulloch and Walter Pitts.
\newblock A logical calculus of the ideas immanent in nervous activity.
\newblock {\em The bulletin of mathematical biophysics}, 5(4):115--133, 1943.
\newblock Publisher: Springer.

\bibitem{gerstein_neuronal_1989}
George~L. Gerstein, Purvis Bedenbaugh, and Ad~MHJ Aertsen.
\newblock Neuronal assemblies.
\newblock {\em IEEE Transactions on Biomedical Engineering}, 36(1):4--14, 1989.
\newblock Publisher: IEEE.

\bibitem{singer_neuronal_1997}
Wolf Singer, Andreas~K. Engel, Andreas~K. Kreiter, Matthias~HJ Munk, Sergio
  Neuenschwander, and Pieter~R. Roelfsema.
\newblock Neuronal assemblies: necessity, signature and detectability.
\newblock {\em Trends in cognitive sciences}, 1(7):252--261, 1997.
\newblock Publisher: Elsevier.

\bibitem{wallace_chasing_2010}
Damian~J. Wallace and Jason~ND Kerr.
\newblock Chasing the cell assembly.
\newblock {\em Current opinion in neurobiology}, 20(3):296--305, 2010.
\newblock Publisher: Elsevier.

\bibitem{palm_cell_2014}
Günther Palm, Andreas Knoblauch, Florian Hauser, and Almut Schüz.
\newblock Cell assemblies in the cerebral cortex.
\newblock {\em Biological cybernetics}, 108(5):559--572, 2014.
\newblock Publisher: Springer.

\bibitem{hutchison_combining_2005}
Rebecca Fay, Ulrich Kaufmann, Andreas Knoblauch, Heiner Markert, and Günther
  Palm.
\newblock Combining {Visual} {Attention}, {Object} {Recognition} and
  {Associative} {Information} {Processing} in a {NeuroBotic} {System}.
\newblock In David Hutchison, Takeo Kanade, Josef Kittler, Jon~M. Kleinberg,
  Friedemann Mattern, John~C. Mitchell, Moni Naor, Oscar Nierstrasz,
  C.~Pandu~Rangan, Bernhard Steffen, Madhu Sudan, Demetri Terzopoulos, Dough
  Tygar, Moshe~Y. Vardi, Gerhard Weikum, Stefan Wermter, Günther Palm, and
  Mark Elshaw, editors, {\em Biomimetic {Neural} {Learning} for {Intelligent}
  {Robots}}, volume 3575, pages 118--143. Springer Berlin Heidelberg, Berlin,
  Heidelberg, 2005.
\newblock Series Title: Lecture Notes in Computer Science.

\bibitem{knoblauch_associative_2005}
Andreas Knoblauch, Heiner Markert, and Guenther Palm.
\newblock An {Associative} {Model} of {Cortical} {Language} and {Action}
  {Processing}.
\newblock In {\em Modeling {Language}, {Cognition} and {Action}}, pages 79--83,
  University of Plymouth, UK, May 2005. World Scientific.

\bibitem{arieli_dynamics_1996}
Amos Arieli, Alexander Sterkin, Amiram Grinvald, and A.~D. Aertsen.
\newblock Dynamics of ongoing activity: explanation of the large variability in
  evoked cortical responses.
\newblock {\em Science}, 273(5283):1868--1871, 1996.
\newblock Publisher: American Association for the Advancement of Science.

\bibitem{markram_blue_2006}
Henry Markram.
\newblock The blue brain project.
\newblock {\em Nature Reviews Neuroscience}, 7(2):153--160, 2006.
\newblock Publisher: Nature Publishing Group.

\bibitem{olshausen_emergence_1996}
Bruno~A. Olshausen and David~J. Field.
\newblock Emergence of simple-cell receptive field properties by learning a
  sparse code for natural images.
\newblock {\em Nature}, 381(6583):607--609, 1996.
\newblock Publisher: Nature Publishing Group.

\bibitem{willmore_characterizing_2001}
Benjamin Willmore and David~J. Tolhurst.
\newblock Characterizing the sparseness of neural codes.
\newblock {\em Network: Computation in Neural Systems}, 12(3):255, 2001.
\newblock Publisher: IOP Publishing.

\bibitem{waydo_sparse_2006}
Stephen Waydo, Alexander Kraskov, Rodrigo~Quian Quiroga, Itzhak Fried, and
  Christof Koch.
\newblock Sparse representation in the human medial temporal lobe.
\newblock {\em Journal of Neuroscience}, 26(40):10232--10234, 2006.
\newblock Publisher: Soc Neuroscience.

\bibitem{quiroga_invariant_2005}
R.~Quian Quiroga, Leila Reddy, Gabriel Kreiman, Christof Koch, and Itzhak
  Fried.
\newblock Invariant visual representation by single neurons in the human brain.
\newblock {\em Nature}, 435(7045):1102--1107, 2005.
\newblock Publisher: Nature Publishing Group.

\bibitem{quiroga_sparse_2008}
R.~Quian Quiroga, Gabriel Kreiman, Christof Koch, and Itzhak Fried.
\newblock Sparse but not ‘grandmother-cell’coding in the medial temporal
  lobe.
\newblock {\em Trends in cognitive sciences}, 12(3):87--91, 2008.
\newblock Publisher: Elsevier.

\bibitem{quiroga_concept_2012}
Rodrigo~Quian Quiroga.
\newblock Concept cells: the building blocks of declarative memory functions.
\newblock {\em Nature Reviews Neuroscience}, 13(8):587--597, 2012.
\newblock Publisher: Nature Publishing Group.

\bibitem{palm_computing_1987}
Gunther Palm.
\newblock Computing with neural networks.
\newblock {\em Science}, 235(4793):1227--1228, 1987.
\newblock Publisher: American Association for the Advancement of Science.

\bibitem{palm_associative_1994}
Günther Palm, Friedhelm Schwenker, and Friedrich~T. Sommer.
\newblock Associative memory networks and sparse similarity preserving codes.
\newblock In {\em From {Statistics} to {Neural} {Networks}}, pages 282--302.
  Springer, 1994.

\bibitem{sa-couto_storing_2020}
Luis Sa-Couto and Andreas Wichert.
\newblock Storing {Object}-{Dependent} {Sparse} {Codes} in a {Willshaw}
  {Associative} {Network}.
\newblock {\em Neural Computation}, 32(1):136--152, January 2020.

\bibitem{sa-couto_attention_2019}
Luis Sa-Couto and Andreas Wichert.
\newblock Attention {Inspired} {Network}: {Steep} learning curve in an
  invariant pattern recognition model.
\newblock {\em Neural Networks}, 114:38--46, June 2019.

\bibitem{desieno_adding_1988}
{DeSieno}.
\newblock Adding a conscience to competitive learning.
\newblock In {\em {IEEE} {International} {Conference} on {Neural} {Networks}},
  pages 117--124 vol.1, San Diego, CA, USA, 1988. IEEE.

\bibitem{lecun_mnist_nodate}
Yann LeCun, Corinna Cortes, and Chris Burges.
\newblock {MNIST} handwritten digit database.

\bibitem{xiao2017/online}
Han Xiao, Kashif Rasul, and Roland Vollgraf.
\newblock Fashion-mnist: a novel image dataset for benchmarking machine
  learning algorithms, 2017.

\bibitem{goodfellow_deep_2016}
Ian Goodfellow, Yoshua Bengio, and Aaron Courville.
\newblock {\em Deep learning}.
\newblock MIT press, 2016.

\bibitem{jang_categorical_2016}
Eric Jang, Shixiang Gu, and Ben Poole.
\newblock Categorical reparameterization with gumbel-softmax.
\newblock {\em arXiv preprint arXiv:1611.01144}, 2016.

\bibitem{mairal_online_2009}
Julien Mairal, Francis Bach, Jean Ponce, and Guillermo Sapiro.
\newblock Online dictionary learning for sparse coding.
\newblock In {\em Proceedings of the 26th {Annual} {International} {Conference}
  on {Machine} {Learning} - {ICML} '09}, pages 1--8, Montreal, Quebec, Canada,
  2009. ACM Press.

\end{thebibliography}


\end{document}